\documentclass[preprint,authoryear]{elsarticle}

\journal{arXiv}
\usepackage{amssymb,amsmath,epsfig}
\usepackage[pagewise]{lineno}
\usepackage{bbm}

\usepackage{setspace}
\usepackage[margin=1in]{geometry}

\DeclareMathOperator{\Exp}{Exp}
\DeclareMathOperator{\Bernoulli}{Bernoulli}

\DeclareMathOperator{\E}{E}

\DeclareMathOperator{\Cov}{Cov}
\DeclareMathOperator*{\argmax}{\arg\!\max}
\def\bs{\boldsymbol}

\makeatletter
\def\ps@pprintTitle{%
\let\@oddhead\@empty
\let\@evenhead\@empty
\def\@oddfoot{\centerline{\thepage}}%
\let\@evenfoot\@oddfoot}
\makeatother

\begin{document}

\title{The SMC' is a highly accurate approximation to the ancestral recombination graph}

\author[har]{Peter R. Wilton\corref{cor}}
\ead{pwilton@fas.harvard.edu}
\author[cu]{Shai Carmi\corref{eq}}
\author[aar]{Asger Hobolth\corref{eq}}

\address[har]{Department of Organismic and Evolutionary Biology, Harvard University, Cambridge, MA, 02138, USA}
\address[cu]{Department of Computer Science, Columbia University, New York, NY, 10027, USA}
\address[aar]{Bioinformatics Research Centre, Aarhus University, Aarhus, Denmark}
\cortext[cor]{Corresponding author}
\cortext[eq]{Contributed equally to this work}

\begin{abstract}
	Two sequentially Markov coalescent models (SMC and SMC') are available as
	tractable approximations to the ancestral recombination graph (ARG). We
	present a Markov process describing coalescence at two fixed points along a
	pair of sequences evolving under the SMC'. Using our Markov process, we
	derive a number of new quantities related to the pairwise SMC', thereby
	analytically quantifying for the first time the similarity between the SMC'
	and ARG. We use our process to show that the joint distribution of pairwise
	coalescence times at recombination sites under the SMC' is the same as it
	is marginally under the ARG, which demonstrates that the SMC' is, in a
	particular well-defined, intuitive sense, the most appropriate first-order
	sequentially Markov approximation to the ARG. Finally, we use these results
	to show that population size estimates under the pairwise SMC are
	asymptotically biased, while under the pairwise SMC' they are approximately
	asymptotically unbiased.
\end{abstract}


\maketitle

\section{Introduction}
\label{sec:introduction}

Of the many models of genetic variation in the field of population genetics,
few have as much relevance in the era of genomics as the ancestral
recombination graph (ARG). The ancestral recombination graph models patterns of
ancestry and genetic variation within sequences experiencing recombination
under neutral conditions \citep{hudson_gene_1991, griffiths_ancestral_1997}.
Under the formulation of \citet{griffiths_ancestral_1997}, lineages recombine
apart and coalesce together back in time to produce a graph structure
describing the ancestral genealogy at each point along a continuous chromosome.
While only a few simple rules govern the process, many aspects of the model are
analytically intractable.

\citet{wiuf_recombination_1999} provided a formulation of the ARG that proceeds
across the chromosome (rather than back in time), producing the genealogy at
each point sequentially. As with the back-in-time formulation of the ARG, at
each point along the chromosome there is a local genealogy describing the
ancestry of the sample at that point, and changes in the genealogy occur at
recombination sites. In this sequential formulation of the ARG, a new lineage
is produced wherever an ancestral recombination event is encountered along the
chromosome.  To produce a new genealogy at the recombination site, the new
lineage is coalesced to the ARG representing the ancestry of all previous
points along the chromosome. This dependence on all previous points makes the
process non-Markovian along the chromosome and (together with a large state
space) makes calculations often intractable.

Approximations to the ARG have been suggested with the goal of modeling
coalescence with recombination in a way that is analytically tractable.
\citet{mcvean_approximating_2005} introduced the \emph{sequentially Markov
coalescent} (SMC). The original formulation of the SMC was sequential,
generating genealogies along the chromosome such that each new genealogy
depends only on the previous genealogy. Like the ARG, the SMC has both a
back-in-time formulation and a sequential formulation. The back-in-time
formulation of the SMC is equivalent to that of the ARG except that coalescence
is allowed only between lineages containing overlapping ancestral material. As
a consequence, in the sequential formulation of the pairwise (${n = 2}$
chromosomes) SMC, each recombination event produces a new pairwise coalescence
time.

\begin{figure}[h!]
	\centering
	\includegraphics[width=1.0\textwidth]{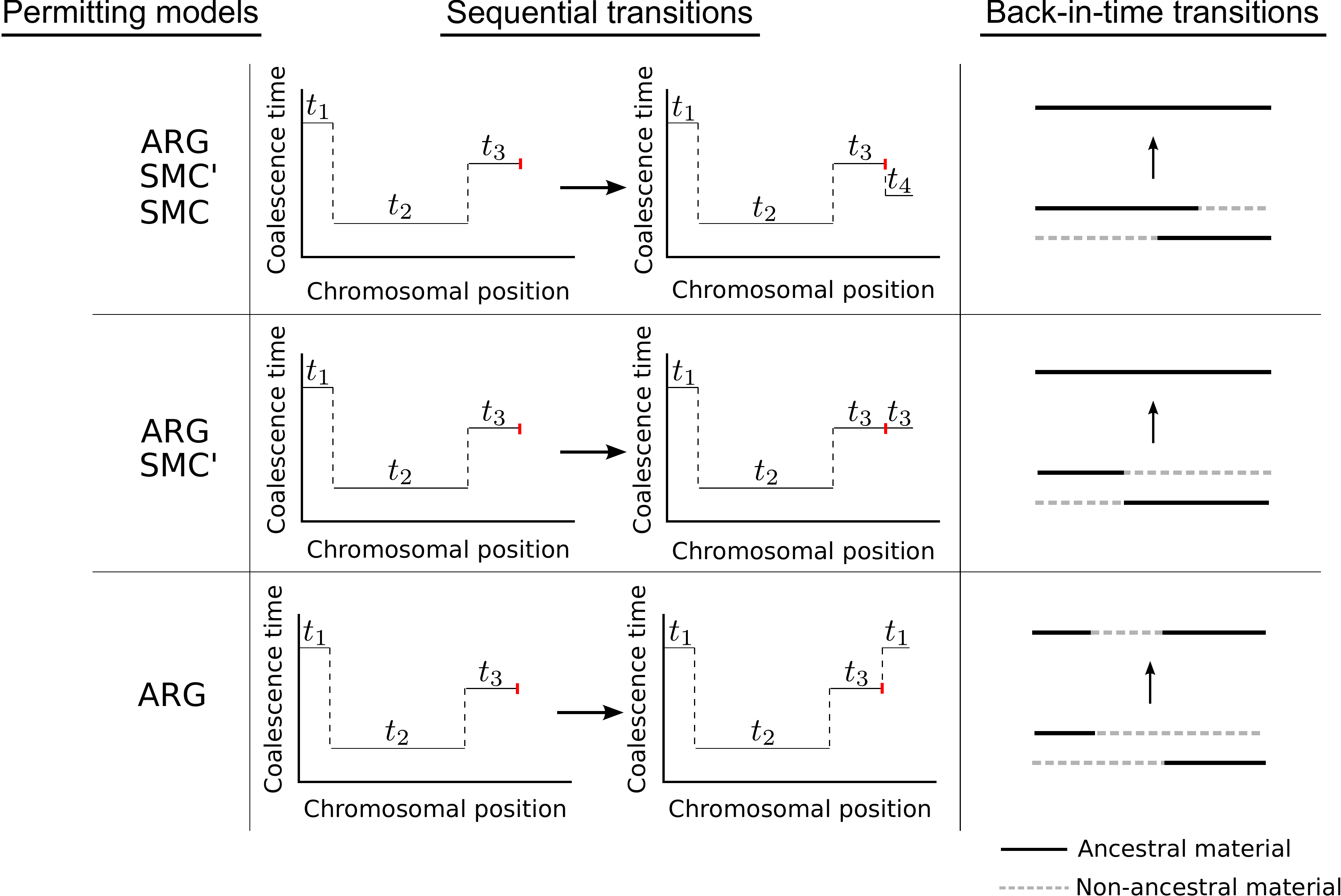}
	\caption{Transitions permitted under the pairwise ARG, SMC', and SMC models.
		Under ``Sequential Transitions,'' a transition occurs left to right
		across the chromosome at the rightmost recombination event (marked with
		red line). The $i$th coalescence time is labeled as $t_i$.  Under
		``Back-in-time transitions,'' the arrow indicates a coalescence event
		that occurs between two aligned ancestral chromosomes, each carrying a
		combination of ancestral (solid black lines) and non-ancestral material
		(dashed gray lines). Ancestral material is defined as a portion of a
		chromosome that is ancestral to the sample.
} 
\label{fig:transitions}
\end{figure}

\citet{marjoram_fast_2006} introduced a slight modification to the SMC, termed
the SMC', which retains the Markov behavior along the chromosome but models
additional coalescence events that make it a closer approximation to the ARG.
Specifically, in the back-in-time formulation of the SMC', coalescence is
allowed between lineages containing either overlapping \emph{or adjacent}
ancestral material. In the sequential formulation of the pairwise SMC', this
means that not every recombination event necessarily produces a new
pairwise coalescence time, since two lineages created by a recombination event can
coalesce back together. Figure~\ref{fig:transitions} shows the transitions that
are permitted under the back-in-time and sequential formulations of the
pairwise ARG, SMC, and SMC'. The sequentially Markov coalescent models have
been used in many recently introduced population-genetic, model-based inference
procedures, including the pairwise SMC (PSMC) model \citep{li_inference_2011},
multiple SMC (MSMC) model \citep{schiffels_inferring_2014}, diCal
\citep{sheehan_estimating_2013}, coalHMM \citep{hobolth_genomic_2007,
dutheil_ancestral_2009}, ARGWeaver \citep{rasmussen_genome-wide_2014}, and in a
study of past human demography based on tracts of identity by state
\citep{harris_inferring_2013}.

The SMC' was shown by simulation to produce measurements of linkage
disequilibrium more similar to the ARG than those produced by the SMC
\citep{marjoram_fast_2006}, and was hence used as the preferred model by some
recent studies \citep{harris_inferring_2013, schiffels_inferring_2014,
zheng_bayesian_2014}. Additionally, a number of recent studies have explored
the theoretical properties of the SMC' \citep{eriksson_sequential_2009,
	harris_inferring_2013, carmi_renewal_2014, schiffels_inferring_2014,
zheng_bayesian_2014}. However, few direct comparisons between the SMC' and the
ARG have been made, and there remain a number of open questions. Here, we show
how the joint distribution of pairwise coalescence times at two fixed points
along a chromosome evolving under the SMC' can be described by a
continuous-time Markov chain. Through analysis of this Markov chain, we
calculate many statistical properties of the pairwise SMC' and compare them to
those of the ARG and SMC. Specifically, for each model of coalescence with
recombination, we compare the following: the joint density ${f_{T_1, T_2}(t_1,
t_2)}$ (Section~\ref{subsec:jointdensity}), the conditional density
${f_{T_2|T_1}(t_2|t_1)}$ (Section~\ref{subsec:conditional}), and the covariance
between $T_1$ and $T_2$, which we show to be equal to the probability that
$T_1$ and $T_2$ are the same (Section~\ref{subsec:probsame}). These quantities
are readily related to measures of linkage disequilibrium in real sequence
data.

Using our two-locus Markov process for the two-locus, pairwise SMC', we also show that the joint
distribution of coalescence times immediately to the left and right of a
recombination event is the same under the SMC' and ARG. This allows us to
calculate the asymptotic bias of the pairwise SMC- and SMC'-based
population-size estimators, which we confirm by simulation. We show that the
SMC' estimator is approximately asymptotically unbiased.

\section{Results}
\label{sec:results}

\subsection{Two-locus Markov chain model for the SMC and SMC'}
\label{subsec:twolocusmodels}

Here, we present back-in-time Markov processes for the two-locus SMC and SMC'.
Previous work has developed analogous two-locus, back-in-time Markov processes
for the ARG. \citet{kaplan_use_1985} first described how the process of
generating coalescence times at two linked loci modeled by the ARG can be
represented as a continuous-time Markov chain, with coalescence and
recombination events causing transitions between states.
\citet{simonsen_markov_1997} explored this process further for the case where
the sample size is $n = 2$ and derived for the ARG many of the quantities we
compare against the SMC' in this paper. Subsequent work has extended this
approach to study two-locus coalescence distributions in the presence of
population structure \citep{eriksson_gene-history_2004} and recurrent
bottlenecks \citep{schaper_linkage_2012}, and to study species-tree concordance
at linked loci \citep{slatkin_concordance_2006} and coalescence histories at
one locus conditional on the history at a nearby locus
\citep{hobolth_markovian_2014}.

\begin{figure}[h!]
	\centering
	\includegraphics[width=0.4\textwidth]{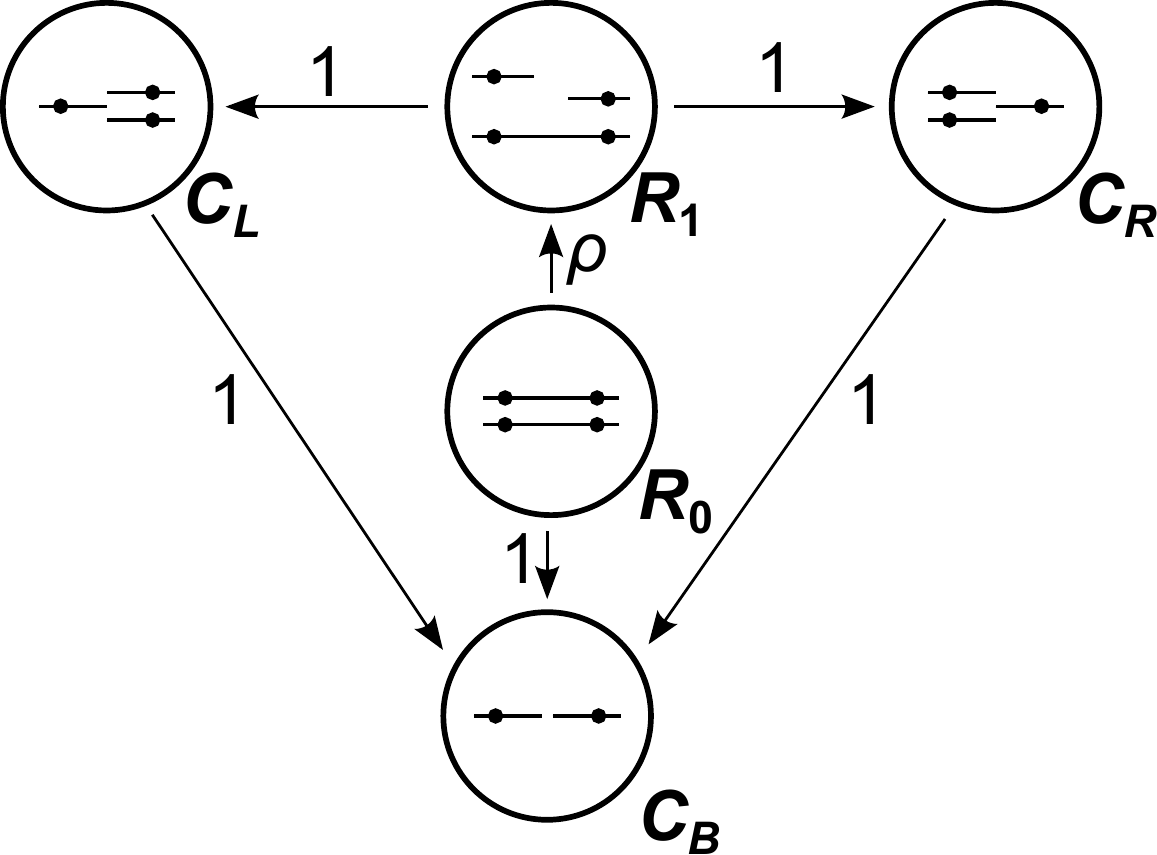}
	\caption{Schematic of the SMC back-in-time Markov process for two loci. The
	process starts in state $\bs{R_0}$, and transitions to other states occur
	with the rates indicated by arrows between states.} 
\label{fig:SMCdiagram}
\end{figure}

We begin by presenting the simpler SMC model, which will provide context for
the more-complex SMC' model. If time is scaled such that the rate of
coalescence is $1$ and the total rate of recombination between the two linked
loci is $\rho/2$, then the two-locus ancestral process under the SMC is the
process depicted in Figure~\ref{fig:SMCdiagram}. The process starts in state
$\bs{R_0}$ with two lineages, each containing linked copies of the two loci.
From $\bs{R_0}$, the process transitions with rate $\rho$ to state $\bs{R_1}$,
in which one of the two chromosomes has experienced a recombination event, or
with rate $1$ to state $\bs{C_B}$, an absorbing state in which both loci have
coalesced. Under the SMC, lineages can only coalesce if they contain
overlapping ancestral material, so after entering $\bs{R_1}$, the process
cannot return to the fully linked state $\bs{R_0}$, and each locus coalesces
independently with rate $1$ from that time onward. Thus, under the SMC, the
joint distribution of coalescence times at two loci is that of 

\begin{equation} 
	(T_1, T_2) \sim (X_0 + RX_L,\ X_0 + RX_R), 
	\label{eq:smcrepresentation}
\end{equation} 
where $X_0 \sim \Exp(1+\rho)$ is the amount of time to leave $\bs{R_0}$, $R
\sim \Bernoulli(\frac{\rho}{1+\rho})$ indicates whether the first event is a
recombination event, and $X_L \sim X_R \sim \Exp(1)$ are the exponential
waiting times until coalescence after the first recombination event. All of
these random variables are independent in the SMC model, so it is
straightforward to calculate many of the quantities we compare in this paper
using this representation.

\begin{figure}[h!]
	\centering
	\includegraphics[width=0.75\textwidth]{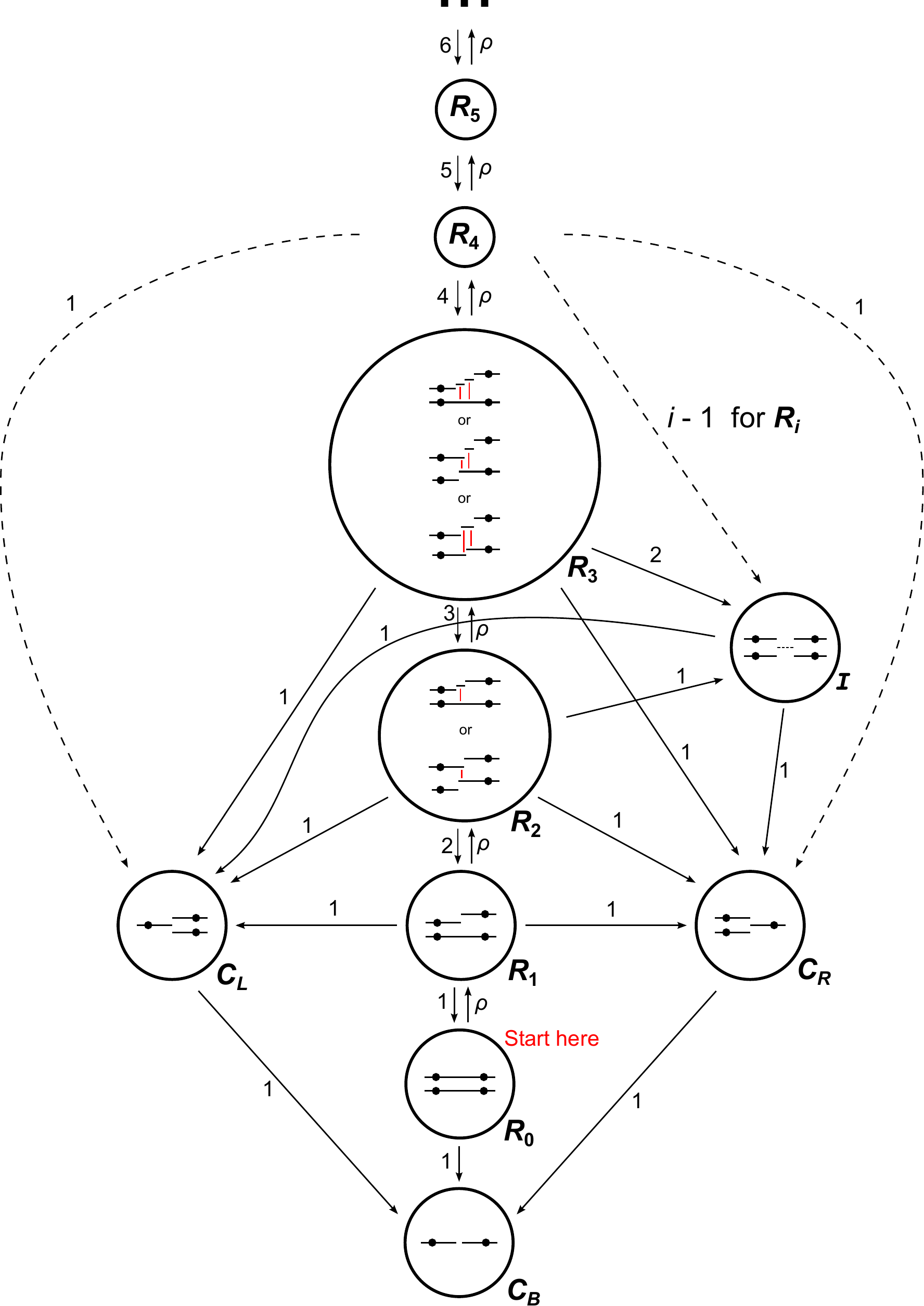}
	\caption{Schematic of the SMC' back-in-time Markov process for two
	loci. Dashed arrows show transition rates that apply for all
	$\bs{R_i}.$ State $\bs{I}$ is the state in which some
	portion of the chromosome between the two focal loci has coalesced but
	neither focal locus has coalesced. The red lines in states $\bs{R_2}$ and
	$\bs{R_3}$ show the coalescence events that take the process to state
	$\bs{I}$.
}
\label{fig:smcprimediagram}
\end{figure}

The defining rule of the SMC' model of coalescence with recombination is that
only ancestral lineages containing overlapping or contiguous ancestral material
can coalesce \citep{marjoram_fast_2006}. The back-in-time process of coalescence
at two fixed loci under this model is the continuous-time Markov chain shown in
Figure~\ref{fig:smcprimediagram}. Under the SMC', it is necessary to model the
number of recombination events that have occurred between the two loci at each
point in time. To see that this is the case, consider the state $\bs{R_2}$ in
Figure~\ref{fig:smcprimediagram}. In this state, two recombination events have
occurred between the focal loci, and neither focal locus has coalesced.
Because lineages can only coalesce to lineages containing overlapping or
adjacent ancestral material, two particular coalescence events would need to
occur before the process returns to state $\bs{R_0}$, regardless of the
placement of the recombination events on the two chromosomes.

The SMC' two-locus Markov process also features an additional state $\bs{I}$,
which is entered when some portion of the chromosome between the focal loci
coalesces before either of the focal loci. Upon entering $\bs{I}$ it becomes
impossible for the process to re-enter the initial, fully-linked state
($\bs{R_0}$), so the remaining times until coalescence at the focal loci become
independent exponential random variables with mean $1$. If $\bs{R_i}$ is the
state in which neither focal locus has coalesced and $i$ recombination events
have occurred between the focal loci, the transition rate into $\bs{I}$ is
$i-1$. This is due to the fact that each recombination event after the first
produces an additional pair of lineages that can coalesce to take the process
to $\bs{I}$. For each state $\bs{R_i}$, $i \ge 1$, the number of lineages that
can coalesce to take the process to $\bs{R_{i-1}}$ is $i$, and the rate of
transitioning to $\bs{R_{i+1}}$ through recombination is $\rho$.  Transitions
to $\bs{C_L}$ and $\bs{C_R}$ occur at rate $1$ whenever the process is in state
$\bs{R_i}$, $i \ge 1$. Following \citet{eriksson_gene-history_2004}, we
disregard any information about linkage between the two loci after one locus
has coalesced, since the rate of coalescence at the uncoalesced locus is $1$
regardless of the state of linkage with the coalesced locus.

For comparison, an analogous two-locus continuous-time Markov chain for the ARG
is presented in Figure~\ref{fig:ARGdiagram}. An equivalent process was studied
by \citet{simonsen_markov_1997} and others. In this model, state $\bs{R_1}$ is
reached when the first event is a recombination event, and state $\bs{R_2}$ is
reached only after a subsequent recombination event occurs on the ancestral
lineage that did not experience the first recombination event, making all
ancestral copies of the two loci unlinked.

\subsection{Joint probability density functions} 
\label{subsec:jointdensity}

Considering the SMC' model above, let $R_0(t)$ represent the probability that
the two-locus ancestral coalescent process is in state $\bs{R_0}$ at time $t$,
and let $R^+(t)$ represent the probability that the process is in any state
$\bs{R_i}$, $i \ge 1$, or state $\bs{I}$, at time $t$. The joint density of
coalescence times at the two focal loci is then

\begin{align}
f_{T_1, T_2}(t_1, t_2) = 
\begin{cases}
	R_0(t_1) & t_1 = t_2 \\
	R^+(t_1)e^{-(t_2-t_1)} & t_1 < t_2 \\
	R^+(t_2)e^{-(t_1-t_2)} & t_1 > t_2,
\end{cases}
\label{eq:jointdistribution}
\end{align}
since $R_0(t)$ is the rate of entering state $\bs{C_B}$ at time $t$, and
$R^+(t)$ is the rate of entering either $\bs{C_L}$ or $\bs{C_R}$ at time
$t$. The joint density for the ARG and SMC is analogously defined, with
$R^+(t)$ representing $\bs{R_1}$ and $\bs{R_2}$ under the ARG and $\bs{R_1}$
under the SMC. For the ARG and the SMC, the number of states is finite and
$R_0(t)$ and $R^+(t)$ can be solved using matrix exponentiation. For the SMC',
there is an infinite number of states, representing the possibility of an
infinite number of recombination events occurring between the two focal loci.
In the Appendix, we derive closed-form expressions for $R_0(t)$, $R^+(t)$, and
$f_{T_1,T_2}(t_1,t_2)$. The main idea in these derivations is to recognize the
structure of the SMC' in Figure~3 as a birth-death process with killing. In
this formulation the states are $\bs{R_i}, \{i \ge 0\}$, a birth corresponds to a
recombination event (and the birth rate is constant), a death corresponds to a
coalescence event (and the death rate is linear), and killing corresponds to
leaving the $\bs{R_i}$ states.

\begin{figure}[h!]
	\centering
	\includegraphics[width=1.0\textwidth]{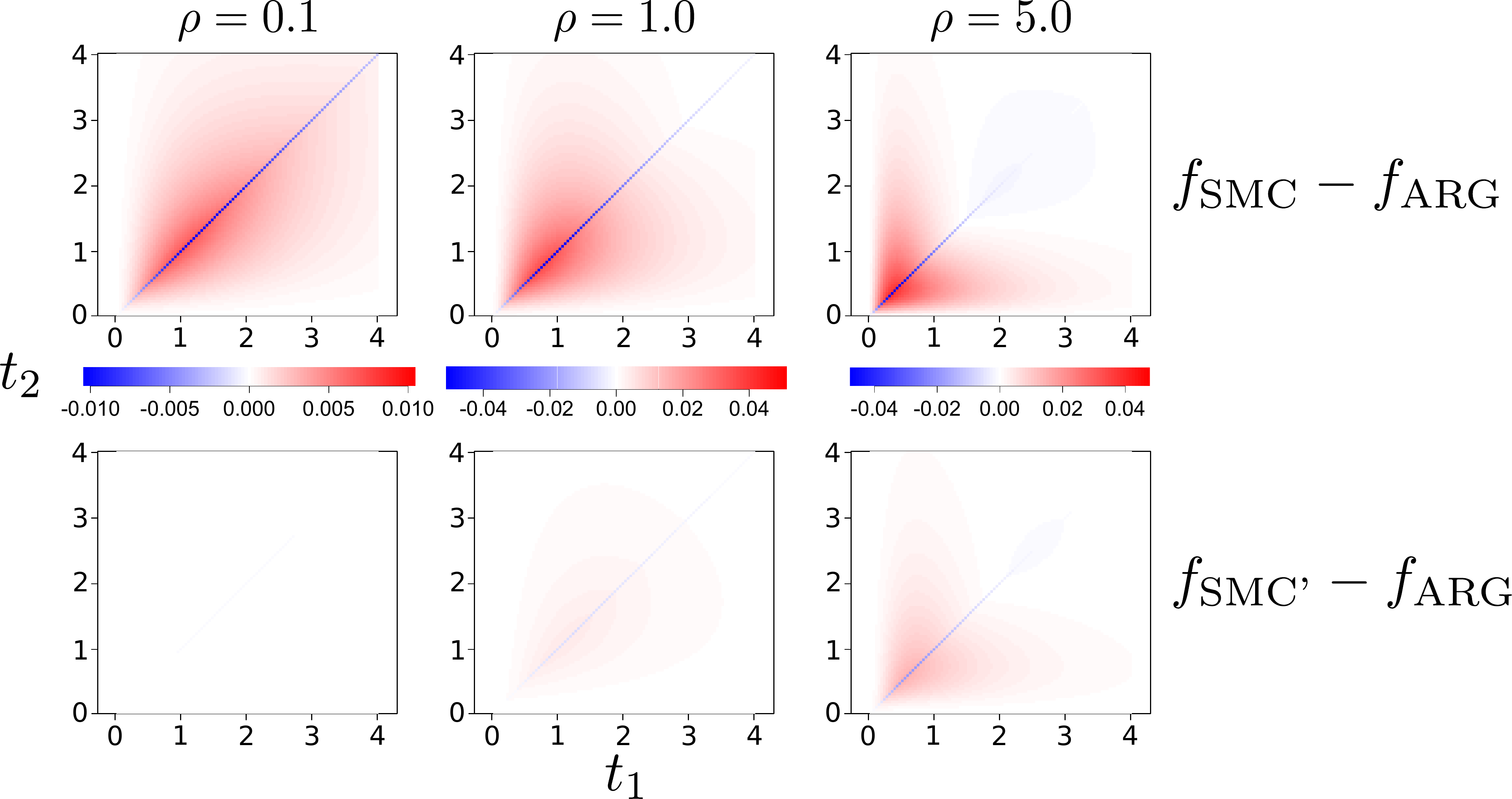}
	\caption{Comparison of the difference in the joint density of coalescence times
		$f_{T_1,T_2}(t_1, t_2)$ between the SMC and ARG (top row) and SMC' and
		ARG (bottom row). Comparisons are made for three different recombination
		rates ($\rho = 0.1, 1.0, 5.0$).}
\label{fig:smc-arg_smcprime-arg_figure}
\end{figure}

Figure~\ref{fig:smc-arg_smcprime-arg_figure} compares the joint coalescence
time distributions under the SMC and SMC', displaying the differences of these
joint distributions with the joint distribution of the ARG. The SMC' provides a
much better fit to the ARG for the range of recombination rates compared. Both
the SMC and the SMC' underestimate the density of outcomes where $T_1 = T_2$,
but this underestimation is substantially less under the SMC'.

\begin{figure}[h!]
	\centering
	\includegraphics[width=0.4\textwidth]{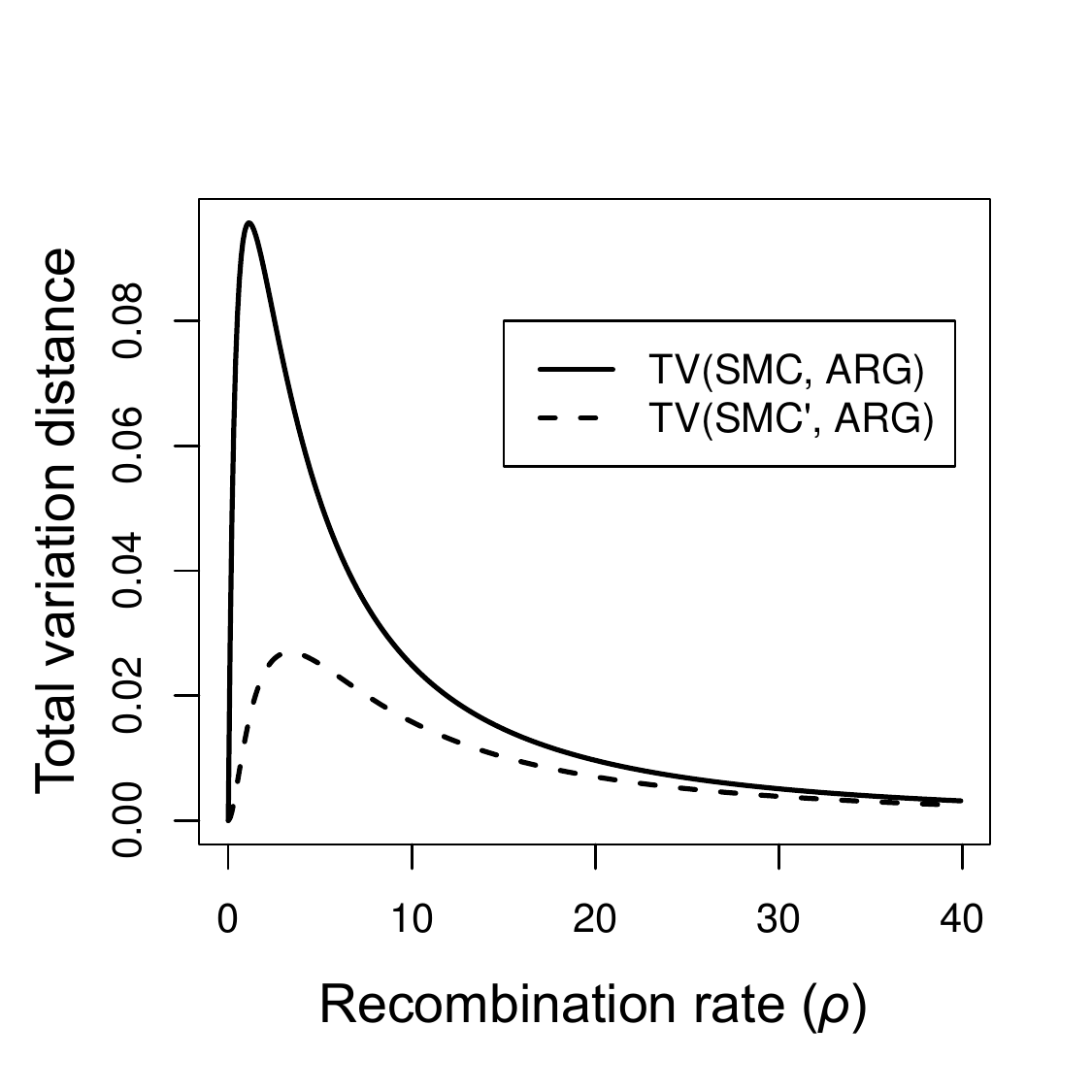}
	\caption{Total variation distance between the SMC and ARG (solid line) and
	the SMC' and ARG (dashed line) as a function of recombination rate. Total
	variation distances were calculated numerically.}
\label{fig:totalvariationdistance}
\end{figure}

To summarize the difference between the joint distributions more succinctly, we
calculated the total variation distance between the SMC and ARG and between the
SMC' and ARG across a range of recombination rates. The total variation
distance between the SMC and the ARG is defined as

\begin{equation}
	TV\left(\textrm{SMC}, \textrm{ARG}\right) = \frac{1}{2}\int_0^\infty\int_0^\infty\left|f^{\textrm{SMC}}(t_1,t_2) - f^{\textrm{ARG}}(t_1,t_2))\right|dt_2dt_1,
\end{equation}
where $f^{\textrm{SMC}}(t_1,t_2)$ and $f^{\textrm{ARG}}(t_1,t_2)$ are the
joint densities $f_{T_1,T_2}(t_1,t_2)$ defined under the SMC and ARG,
respectively. The total variation distance between the SMC' and ARG is
similarly defined. Figure~\ref{fig:totalvariationdistance} shows the total
variation distance from the ARG for the SMC and SMC' over a range of
recombination rates. Total variation distances were calculated numerically. For
both the SMC and SMC', the total variation distance was maximized at some
intermediate recombination rate, approximately $\rho = 1.1$ for the SMC and
$\rho = 3.2$ for the SMC'.

\subsection{Conditional distribution of coalescence times}
\label{subsec:conditional}

In this section we consider the distribution of coalescence times at one locus
given the coalescence time at the other. The conditional density of $T_2$ given
$T_1$, $f_{T_2|T_1}(t_2|t_1)$, can be calculated by dividing the joint
density by the marginal distribution of coalescence times at the left locus:

\begin{equation}
	f_{T_2|T_1}(t_2|t_1) = \frac{f_{T_1, T_2}(t_1, t_2)}{e^{-t_1}}.
\label{eq:conditionaldistribution}
\end{equation}

\begin{figure}[h!]
	\centering
	\includegraphics[width=1.0\textwidth]{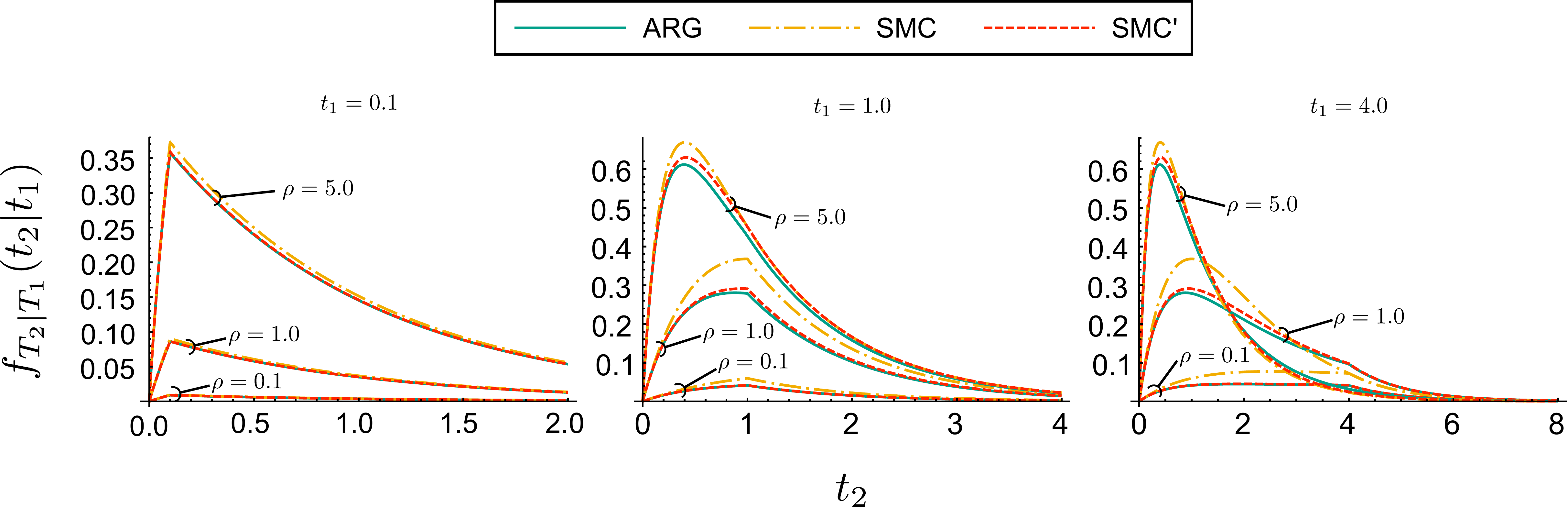}
	\caption{Comparison of densities of coalescence times $t_2$ at the right locus
	conditioned upon coalescence times $t_1$ at the left locus. Conditional
	densities $f_{T_2|T_1}(t_2|t_1)$ are shown for the ARG, SMC, and SMC' models for
	three different rates of recombination between the two loci ($\rho =
	0.1, 1.0, 5.0$) and three different conditioned-upon coalescence times $t_1$
	at the left locus ($t_1 = 0.1, 1.0, 4.0$). The area under each curve
	is $P(T_2 \neq t_1 | T_1 = t_1)$; the conditional probabilities $P(T_2 =
	t_1 | T_1 = t_1)$ are not shown.
}
\label{fig:conditional_density_comparison}
\end{figure}

\citet{hobolth_markovian_2014} introduced a framework for modeling the
distribution of $T_2$ given $T_1$ using a time-inhomogeneous continuous-time
Markov chain. (Note that the model called SMC' in
\citet{hobolth_markovian_2014} is an SMC'-like model of two loci that is not
based on the continuous-chromosome SMC'. It is different from the SMC' model we
consider here.) This framework can be extended to the SMC', producing the
continuous-time Markov chain shown in
Figure~\ref{fig:smcprime_conditional_diagram}.
Figure~\ref{fig:conditional_density_comparison} compares the conditional
density $f_{T_2|T_1}(t_2|t_1)$ of coalescence times $t_2$ at the right locus
conditioned upon the coalescence times $t_1$ at the left locus for different
values of $t_1$ and $\rho$.

We note that recently it was proposed that the mutation rate could be estimated
by simulation-based calibration of the increase in mean heterozygosity when
moving away from a site of known, low heterozygosity
\citep{lipson_calibrating_2015}. Our expressions for the conditional
distribution of coalescence times could provide theoretical expectations for
such a statistic.

\subsection{Probability of coalescence times being equal to covariance between
coalescence times}
\label{subsec:probsame}

In the two-locus, back-in-time Markov processes for the SMC, SMC', and ARG,
$T_1$ and $T_2$ are equal when the state $\bs{C_B}$ is entered through
$\bs{R_0}$ rather than $\bs{C_L}$ or $\bs{C_R}$. For the ARG,
\citet{simonsen_markov_1997} showed that the probability that $T_1$ is equal to
$T_2$ is 

\begin{equation}
	P_{\textrm{ARG}}(T_1 = T_2) = \frac{\rho+18}{\rho^2 + 13\rho + 18}.
\label{eq:psamearg}
\end{equation}
Under the SMC \citep{mcvean_approximating_2005},

\begin{equation}
	P_{\textrm{SMC}}(T_1 = T_2) = \frac{1}{1+\rho}.
\label{eq:psamesmc}
\end{equation}
\citet{eriksson_sequential_2009} used the sequential formulation of the SMC' to
show that

\begin{align}
\begin{split}
	P_{\textrm{SMC'}}(T_1 = T_2) &= \int_0^\infty e^{-t}e^{-\rho \lambda(t)}dt \\
	     &= 2^{\rho/2} e^{-\rho/4} (-\rho )^{-\frac{1}{2}-\frac{\rho }{4}}
	\left[\Gamma\left(\frac{2+\rho}{4}\right)-
	\Gamma\left(\frac{2+\rho }{4},-\frac{\rho}{4}\right)\right],
\end{split}
\label{eq:psamesmcprime}
\end{align}
where ${\lambda(t) = \frac{1}{4}\left(1-e^{-2 t}+2t\right)}$ is the exponential
rate of encountering a change in coalescence time when the local coalescence
time is $t$ and $\Gamma(a, b) = \int_b^\infty x^{a-1}e^{-x}dx$ is the
incomplete gamma function.


For the ARG and SMC, the covariance $\Cov[T_1,T_2]$ is equal to $P(T_1=T_2)$.
\citet{eriksson_sequential_2009} showed by simulation that this is also true of
the SMC'. Here we present a short proof that this is the case for any two-locus
model of coalescence where the marginal distribution of coalescence times is
exponential with rate~$1$. 

The expectation $\E[T_1T_2]$ can be derived using the fact that ${(a-b)^2 =
a^2 + b^2 - 2ab}$:

\begin{align}
	\begin{split}
	2\E[T_1T_2] &= \E[T_1^2] + \E[T_2^2] - \E[(T_1-T_2)^2] \\
	            &= 2 + 2 - \E\big[(T_1-T_2)^2|T_1 \neq T_2\big]P(T_1 \neq T_2) \\
	            &= 4 - 2P(T_1 \neq T_2).
	\end{split}
	\label{eq:shortproof}
\end{align}
The final equality in \eqref{eq:shortproof} follows from the fact that
$|T_1-T_2|$ has an Exponential distribution with rate $1$ when $T_1 \neq T_2$.
Therefore $\E[T_1T_2] = 2 - P(T_1 \neq T_2)$ and 

\begin{align}
	\begin{split}
		\Cov[T_1,T_2] &= \E[T_1T_2]-\E[T_1]\E[T_2]\\
		              &=\E[T_1T_2]-1\\
					  &= P(T_1 = T_2).
	\end{split}
\end{align}

This result holds in other situations with exponential coalescence times, for
example in the context of the population-divergence model considered by
\citet{eriksson_sequential_2009} (in which case the marginal distribution is
exponential plus a constant) and for the various covariances used by
\citet{mcvean_genealogical_2002} to calculate $\sigma_d^2$, the approximation
to the linkage disequilibrium measure $r^2$.

It is interesting to consider $\Cov[T_1,T_2] = P(T_1 = T_2)$ when $\rho$ is
small. For the ARG, consideration of \eqref{eq:psamearg} shows that
${\Cov[T_1,T_2] = P_{\textrm{ARG}}(T_1 = T_2) = 1-2\rho/3 + O(\rho^2)}$.
Likewise, for the SMC, \eqref{eq:psamesmc} shows that ${\Cov[T_1, T_2] =
P_{\textrm{SMC}}(T_1 = T_2) = 1- \rho + O(\rho^2)}$. For the SMC', the integral
representation of $P_{\textrm{SMC'}}(T_1 = T_2)$ in \eqref{eq:psamesmcprime}
allows for the calculation of this quantity as a first-order expansion in $\rho$:

\begin{align}
\begin{split}
	\Cov[T_1, T_2] = P_{\textrm{SMC'}}(T_1 = T_2) &= \int_0^\infty e^{-t}e^{-\rho \lambda(t)}dt \\
		&= 1 - \rho \int_0^\infty e^{-t} \lambda(t) dt + O(\rho^2) = 1-\frac{2\rho}{3} + O(\rho^2).
\end{split}
\label{eq:psamesmcprimeexpansion}
\end{align}
Thus, $\Cov[T_1,T_2]$ (or $P(T_1 = T_2)$) is the same up to order $\rho^2$
under the ARG and SMC'.

\subsection{Coalescence times at recombination sites}

In this section, we show that the joint distribution of coalescence times on
either side of a recombination event is the same under the SMC' and marginally
under the ARG, and we derive this distribution. Consider the continuous-time
Markov chains representing the two-locus SMC' and ARG models
(Figures~\ref{fig:smcprimediagram} and \ref{fig:ARGdiagram}, respectively) in
the limit of $\rho \to 0$ and conditioning on the first event being a
recombination event. These processes represent the joint distribution of
coalescence times on either side of a recombination event under the ARG and
SMC'. In both of these processes, the waiting time until the first event,
conditional on that event being a recombination event, has an exponential
distribution with rate $1+\rho$, which converges to $1$ as $\rho \to 0$. After
that first recombination event, the rate of all additional recombination events
converges to zero in the $\rho \to 0$ limit, so all of the remaining events
must be coalescence events, each of which occurs with rate $1$. Under the ARG
and the SMC', the coalescence events that are possible from state $\bs{R_1}$
are the same. Thus, the joint distribution of coalescence times at
recombination sites is the same under the SMC' and the ARG.

\begin{figure}[h!]
	\centering
	\includegraphics[width=0.8\textwidth]{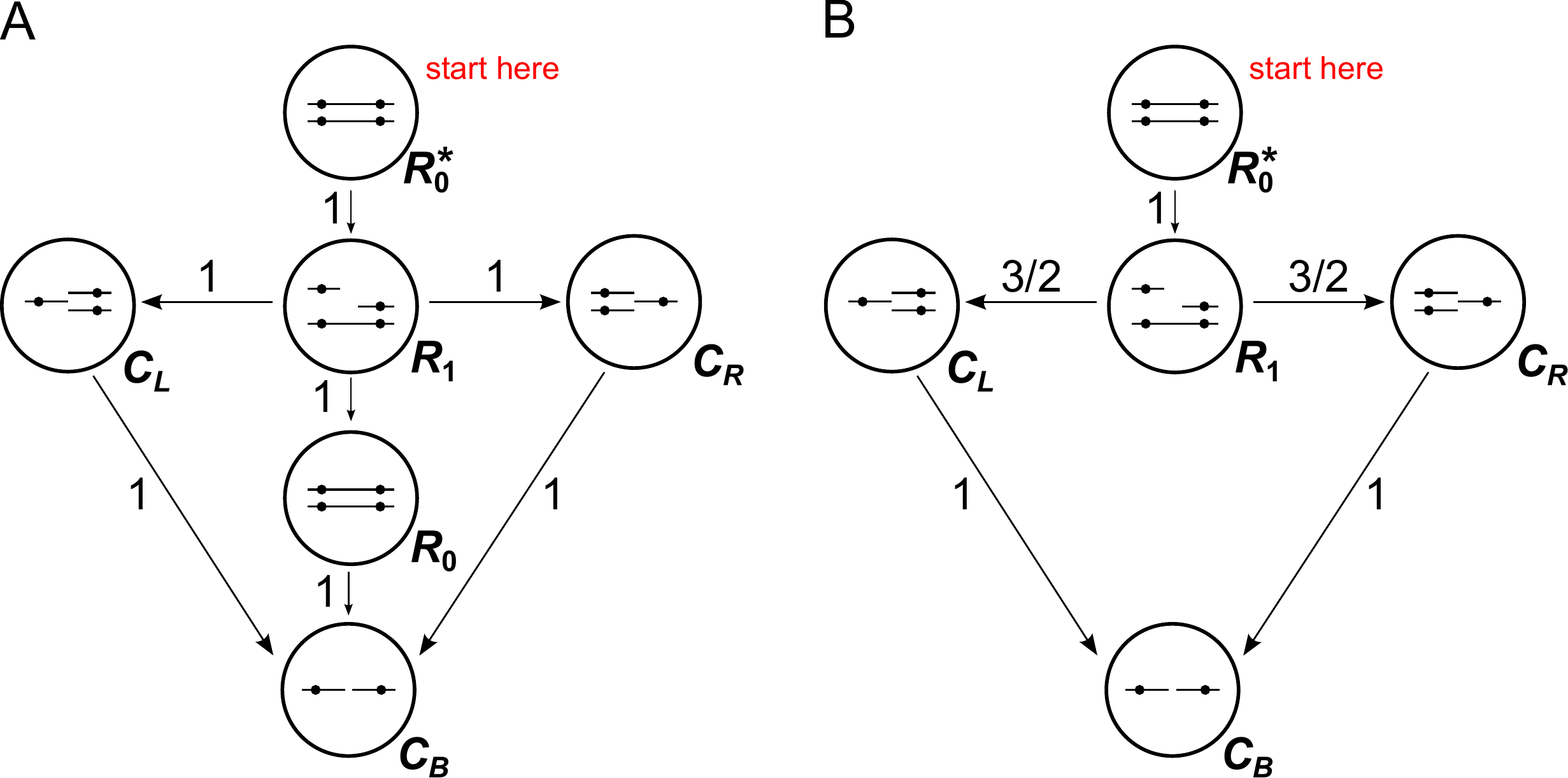}
	\caption{Two-locus continuous-time Markov chains representing the ARG and
		SMC' models in the $\rho \to 0$ limit, conditional on the first event
		being a recombination event. These processes represent the joint
		distribution of coalescence times on either side of a recombination
		site under the ARG and SMC'. The state $\bs{R_0^*}$ is a special starting
		state out of which the first event is always a recombination event.
		Panel A shows the process unconditional on whether $S=T$, and Panel B
		shows the process conditional on $S \neq T$. The model representing the
		joint distribution of coalescence times at recombination sites under
		the SMC is equivalent to the model in Panel B with the transition rates
		from $\bs{R_1}$ to $\bs{C_L}$ and $\bs{C_R}$ equal to $1$ instead of
	$3/2$.}
\label{fig:ARGSMCPrimeConditional}
\end{figure}

Figure~\ref{fig:ARGSMCPrimeConditional}A shows the two-locus continuous-time
Markov chain representing this conditional process. This Markov chain starts in
a special initial state $\bs{R_0^*}$, out of which the first event is always a
recombination event, which happens with rate $1$, as described above. In
previous sections, we used $T_1$ and $T_2$ to represent the coalescence times
at two loci some fixed distance apart. To avoid confusion, in this section we
use $S$ and $T$ to represent the coalescence times on the left and right sides
of a recombination event, respectively.

Recombination events are visible in sequence data only if they change the local
coalescence time. Thus, it is of special interest to condition on $S \neq T$ in
the above model in order to derive the joint distribution of coalescence times
on either side of a change in coalescence times under the ARG and SMC'.
Conditioning on $S \neq T$, the transition out of $\bs{R_1}$ must be into
either $\bs{C_L}$ or $\bs{C_R}$. These transitions occur with conditional rate
$3/2$, since the total rate of leaving $\bs{R_1}$ is three in the unconditional
model, and two of the ways of leaving $\bs{R_1}$ result in the coalescence
times being different.

The continuous-time Markov chain representing coalescence times on either side
of a change in coalescence times (i.e., at recombination sites where $S \neq T$)
is shown in Figure~\ref{fig:ARGSMCPrimeConditional}B. Under this model, the
joint distribution of $S$ and $T$ is that of

\begin{equation}
	(S, T) \sim \big(X_1 + X_2 + RX_3,\ X_1 + X_2 + (1-R)X_3\big),
	\label{eq:smcprimejointrepresentation}
\end{equation}
where $X_1 \sim \Exp(1)$, $X_2 \sim \Exp(3)$, $R \sim \Bernoulli(1/2)$, $X_3
\sim \Exp(1)$, and the random variables are independently distributed.

Under the SMC, the continuous-time Markov chain representing the joint
distribution of coalescence times at recombination sites is equivalent to the
model in Figure~\ref{fig:ARGSMCPrimeConditional}B with the transition rates
from $\bs{R_1}$ to $\bs{C_L}$ and $\bs{C_R}$ equal to $1$ instead of $3/2$.
Under this model for the SMC, the joint distribution of coalescence times on
either side of a recombination event is that of 

\begin{equation}
	(S,T) \sim (X_1+X_2, X_1+X_3),
\end{equation}
where $X_1$, $X_2$, and $X_3$ are all mutually independent exponential random
variables with rate $1$. 

In the Supporting Information, we use these Markov processes to derive the
joint, marginal, and conditional distributions of coalescence times at
recombination sites under the ARG, SMC', and SMC. These calculations confirm
previous derivations of \citet{carmi_renewal_2014} for the SMC' and
\citet{li_inference_2011} for the SMC.

\subsubsection{SMC' as the canonical first-order Markov approximation to ARG}
Under the sequential formulation of the continuous-chromosome ARG, SMC, and
SMC' models, the infinitesimal probability of a recombination event occurring
in the interval $(x, x+dx)$ given the coalescence time $s$ at $x$ is $s\,dx$.
This fact, together with the fact that the joint distribution of coalescence
times at recombination sites is the same under the ARG and SMC' (whether or not
the coalescence time changes), implies that the conditional distribution of
coalescence times at point $x+dx$ given the coalescence time at point $x$ is
the same under the SMC' and ARG. 

This fact demonstrates that the pairwise SMC' is the canonical first-order
Markov approximation for the pairwise ARG. Given an infinite-order Markov chain
$\{X_i, i = 0, 1, 2, \dots\}$, where the distribution of each $X_j$ depends on
all previous $X_i$, $i < j$, the canonical $k$-order Markov approximation to
$\{X_i\}$ is the Markov chain $\{X_i^{[k]}\}$ satisfying 

\begin{equation}
	P(X_n^{[k]}|X_{n-1}^{[k]} = x_{n-1}, \hdots, X_{n-k}^{[k]} = x_{n-k}) =
	P(X_n|X_{n-1} = x_{n-1}, \hdots , X_{n-k} = x_{n-k}).
\label{eq:kordercanonical}
\end{equation}
That is, the transition probabilities under the $k$-order canonical Markov
approximation are equal to the transition probabilities conditional on the
previous $k$ states under the infinite-order chain. See
\citet{schwarz_noninvariance_1976}, \citet{fernandez_markov_2002}, and
\citet{gallo_markov_2013} for examples of mathematical studies of canonical
Markov approximations of infinite-order Markov chains.

Here we informally extend the terminology of canonical Markov approximations to
continuous processes. The SMC' is the canonical first-order Markov
approximation to the ARG because the distribution of coalescence times at
$x+dx$ conditional on the coalescence time at $x$ is the same under the ARG (an
infinite-order, sequentially non-Markovian continuous process) and the SMC' (a
first-order sequentially Markov continuous process). In this sense, the
SMC' is the most natural first-order sequentially Markov approximation to the
ARG.

\subsection{Asymptotic bias of the population-size estimators under SMC and SMC'}

Given the joint density of pairwise coalescence times at recombination sites
under the ARG, it is possible to determine the asymptotic bias of
maximum-likelihood population size estimators derived from the pairwise SMC
and SMC' likelihood functions. These likelihood functions give the probability
of observing a sequence of pairwise coalescence times and corresponding segment
lengths across a chromosome under the SMC and SMC' models. Related likelihood
functions (allowing for variable historical population size) are implicitly
maximized in the PSMC and MSMC inference procedures
\citep[][respectively]{li_inference_2011, schiffels_inferring_2014}. These
inference procedures are hidden Markov model (HMM) methods in which the local
coalescence times (or genealogies) and segment lengths are hidden states
inferred from sequence data. 

Here, we consider the estimators that would be obtained if the hidden states in
these models were actually observable \citep[see also][]{kim_can_2014}. We are
motivated by the fact that any biases of the estimators we investigate are
likely to be inherent in the full HMM-based inference procedures, since these
biases would be present even with perfect knowledge of an infinite number of
coalescence times. Furthermore, by analyzing estimators derived from the hidden
coalescence states, we isolate the bias that is due to choice of coalescent
algorithm (SMC vs.\ SMC') from the bias due to the mutation model or
discretization of the continuous hidden states in a full HMM approach to
inference.

To investigate the asymptotic properties of these estimators, we assume that
data are generated under the ARG, such that at a fixed point the distribution
of pairwise coalescence times is exponential with rate equal to $1$ and an
ancestral segment of length $l$ recombines back in time at rate $\rho l/2$.
Segment lengths are measured in units of the true scaled recombination
parameter $\rho$. Data generated under this model can be represented as a
sequence of pairwise coalescence times and corresponding segment lengths:
$\{(t_i, l_i): 1 \le i \le k\}$. 

We are interested in estimating a single relative population size $\eta$
(defined relative to the true population size, $N$). If the data are modeled by
the SMC or SMC', the likelihood of a particular value of $\eta$ is

\begin{equation}
	L(\eta | \{(t_i, l_i)\}) = \frac{1}{\eta}e^{-\frac{t_1}{\eta}} \prod_{i=2}^k q(t_{i}|t_{i-1};\eta) \prod_{i=1}^k \lambda(t_i;\eta) e^{-\lambda(t_i;\eta) l_i},
	\label{eq:fulllikelihood}
\end{equation}
where $q(t|s)$ is the transition function and $\lambda(t;\eta)$ is the rate
of encountering the end of a segment given $t$, with both quantities
pertaining to the sequentially Markov coalescent model being used to calculate
the likelihood.

In the Appendix, we show that if the SMC is used, the maximum-likelihood
estimate of $\eta$ converges to approximately $0.95$ as the chromosome gets
infinitely long.  If the SMC' is used, the estimate is approximately unbiased
in the same limit. If the data are reduced to just the segment ages, the
likelihood equation is

\begin{equation}
	L(\eta | \{(t_i, l_i)\}) = \frac{1}{\eta}e^{-\frac{t_1}{\eta}} \prod_{i=2}^k q(t_{i}|t_{i-1};\eta).
	\label{eq:reducedlikelihood}
\end{equation}
Using this reduced likelihood, the asymptotic maximum-likelihood estimate is
asymptotically unbiased under the SMC'. Under the SMC, the reduced likelihood
and the full likelihood produce the same maximum-likelihood estimate (see
Appendix).

\begin{figure}[h!]
	\centering
	\includegraphics[width=0.5\textwidth]{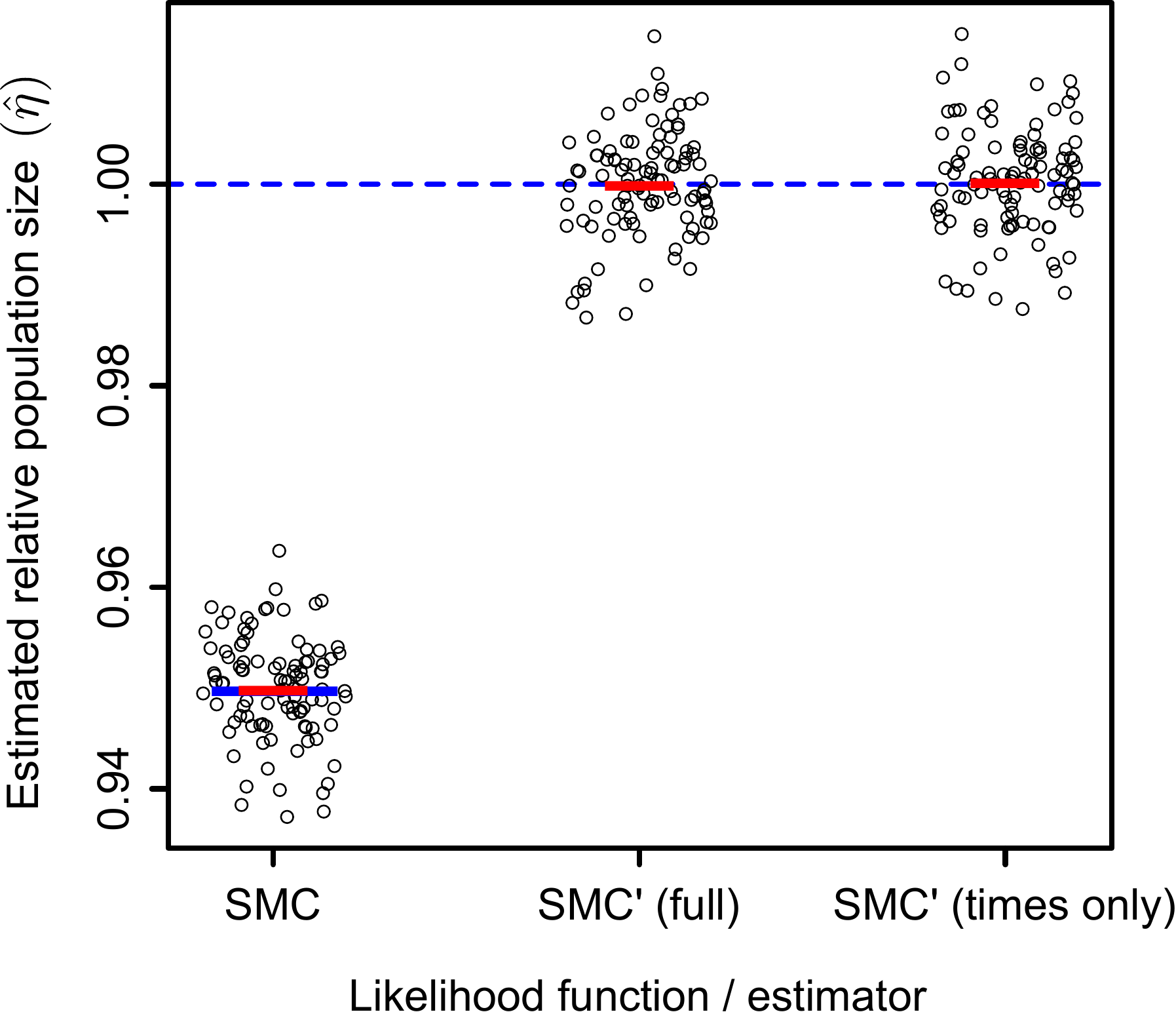}
	\caption{Maximum-likelihood estimates of relative population size with three
		different Markov chain likelihood functions. For each simulation, the
		segment lengths and coalescence times were taken from 100 independent
		pairs of chromosomes, with each chromosome being of length $\rho = 4Nr =
		1000$ simulated under the ARG. A maximum-likelihood estimate was
		calculated using the SMC, SMC', and times-only SMC' likelihood functions
		(equations \eqref{eq:smclikelihood}, \eqref{eq:smcprimelikelihood}, and
		\eqref{eq:reducedlikelihood}, respectively). The true scaled
		population size is $\eta = 1$, shown with the dashed blue line. The
		predicted asymptotic bias of the SMC likelihood function ($\hat{\eta}
	\approx 0.95$) is shown with a solid blue line. The sample mean of the
estimates calculated with each likelihood function is shown with a solid red
line. A total of $100$ simulated datasets were analyzed.}
\label{fig:inferencesimulations}
\end{figure}

We confirm the asymptotic bias of the SMC estimator and the apparent lack of
asymptotic bias of the SMC' estimators by simulation.
Figure~\ref{fig:inferencesimulations} shows 100 simulated estimates calculated
using the SMC, SMC', and reduced SMC' likelihood functions. Each estimate was
calculated using 100 independent pairs of chromosomes simulated under the ARG,
with each chromosome of total length $4Nr = 1000$, where $N$ is the diploid
size and $r$ is the per-generation probability of recombination.  Likelihood
functions were multiplied across independent pairs of chromosomes, and the same
set of simulations was used to produce the estimates for all three likelihood
functions.

\section{Discussion}

We have presented a continuous-time Markov chain that describes the pairwise
coalescence times at two fixed loci evolving under the SMC' model of
coalescence with recombination. We analyzed this Markov chain to derive
the joint distribution of coalescence times at the two loci and the conditional
distribution of coalescence times at one locus given the coalescence time at
the other. We compared these distributions to those of the ARG and SMC
models and found that the difference between the ARG and the SMC' was much less
than the difference between the ARG and the SMC. 

We showed that the conditional distribution of coalescence times at point
$x+dx$ given the coalescence time at $x$ is the same under the ARG and SMC'.
This implies that the SMC' is the canonical first-order approximation to the
pairwise ARG. However, this correspondence is true only of the
continuous-chromosome models. If instead the ARG is a model of the genealogies
at a sequence of discrete loci, then the first-order canonical Markov
approximation is the Markov approximation obtained by modeling a conditional
ARG between every successive pair of loci. This model was studied by
\citet{hobolth_markovian_2014}, who referred to the model as a ``natural''
Markov approximation to the ARG. Conceptually similar sequentially Markov
coalescent models have been used in the so-called ``coalescent hidden Markov
model'' family of methods \citep{hobolth_genomic_2007, dutheil_ancestral_2009,
mailund_estimating_2011}.

\citet{chen_fast_2009} presented a method of simulating data under higher-order
sequentially Markov approximations to the ARG, where the ARG of some number of
preceding loci is retained in the process of generating the marginal genealogy
at a given locus. They showed by simulation that higher-order approximations
generate times until most recent common ancestry that are more consistent with
the ARG than do lower-order approximations, but little theoretical work on
these higher-order Markov approximations has been done.

The two-locus Markov chains analyzed in this paper assume a single well-mixed
population, but natural populations often have more complex demographic
histories, featuring, for example, variable historical population sizes,
migration between subpopulations, and/or past divergence from other
populations. The theoretical properties of the sequential,
across-the-chromosome formulations of the pairwise SMC and SMC' with variable
population sizes have been studied previously \citep{li_inference_2011,
schiffels_inferring_2014}. \citet{eriksson_sequential_2009} used simulation to
study two-locus properties of the SMC' with population bottlenecks, migration
between subpopulations, and divergence between populations. They found that the
SMC' generally performs well in these contexts. The two-locus Markov chains we
study here could be extended to include these features (as was done for the ARG
by \citet{lessard_two-locus_2003} and \citet{eriksson_gene-history_2004}),
which would provide a framework for calculating exact quantities for the
two-locus SMC and SMC' in the context of structured populations. We leave this
for future work.

We calculated the asymptotic bias of a population size estimator under the
pairwise SMC to be approximately 95\% of the true population size. This is not
a very large bias, but given the
continued use of the SMC model in population-genomic inference methods
\citep{palamara_length_2012, sheehan_estimating_2013,
rasmussen_genome-wide_2014}, there is an apparent need to re-examine the
consequences of using the simpler SMC model instead of the slightly more
complicated SMC' model. For example, it will be important to consider whether
including the possibility of varying population sizes will increase or
decrease asymptotic bias. In this context, using the SMC as a basis for a
likelihood function may also bias the estimates of the magnitude and timing of
population size changes, since the longer segments produced by the ARG will
seem younger when they are modeled under the SMC. 

Depending on the particular application, it may sometimes be mathematically
difficult to employ the SMC' instead of the SMC. Nevertheless, the SMC' is the
model underlying two recently introduced population-genetic inference methods:
the multiple SMC (MSMC) method of \citet{schiffels_inferring_2014} (which
simplifies to a PSMC' inference procedure when the number of haplotypes is two)
and a procedure based on the distribution of distances between heterozygous
bases, introduced by \citet{harris_inferring_2013}. In each case it was
acknowledged that the SMC' provided more accurate results than the SMC.

From the arguments that led to the development of the continuous-time Markov
chains representing the joint distribution of coalescence times at
recombination sites (Figure~\ref{fig:ARGSMCPrimeConditional}), it seems that
the joint distribution of coalescence times on either side of a recombination
event will be the same under a variety of demographic scenarios. If one were to
allow variable population historical size, the waiting time until the
conditioned-upon recombination event would still be the same under the SMC' and
ARG, and the remaining coalescence events would also be distributed
identically. Likewise, when there is population substructure with migration
between subpopulations, the distribution of events occurring at recombination
sites should be the same under the SMC' and ARG. Finally, when there are more
than two haplotypes sampled, it seems that the joint distributions of
genealogies on either side of a recombination event would be the same between
the SMC' and the ARG marginally. These ideas need to be properly explored in
future studies, but they suggest that asymptotic bias due to using the SMC' in
place of the ARG will be minimal under a variety of demographic scenarios.

\section{Acknowledgments}
We thank Erik van Doorn and S{\o}ren Asmussen for identifying the
correspondence to birth-death models with killing (see Appendix). We are
grateful to John Wakeley and Paul Marjoram, and two anonymous reviewers for
comments that helped improve this article. S.C. thanks the Human Frontier
Science Program for financial support.

\section{References}
\label{sec:references}
\bibliographystyle{genetics}
\bibliography{smcprime_paper}

\section{Appendix}

\subsection{Derivation of joint density of pairwise coalescence times at two
loci}

To calculate the joint density of coalescence times, it is necessary to
calculate $R_j(t)$, the probability that the SMC' two-locus Markov process
(Figure\ \ref{fig:smcprimediagram}) is in state $\bs{R_j}$ at time $t$, and
$I(t)$, the probability that the SMC' process is in state $\bs{I}$ at time $t$.
To solve for $R_j(t)$, one can use the forward Kolmogorov equation (for $j \ge
1$)

\begin{equation}
	R_j'(t) = \rho R_{j-1}(t) + (j+1)R_{j+1}(t) - (2j+1+\rho)R_j(t).
	\label{eq:forwardkolmogorov}
\end{equation}
Through substitution, the solution to \eqref{eq:forwardkolmogorov} can be shown
to be

\begin{equation}
	R_j(t) = R_0(t)\frac{\left[\frac{\rho}{2}(1-e^{-2t})\right]^j}{j!}.
	\label{eq:smcprimesolution1}
\end{equation}
To find $R_0(t)$, we note that it is equal to $f_{T_1,T_2}(t, t)$ (see
Eq.~\eqref{eq:jointdistribution}). In turn, ${f_{T_1,T_2}(t, t) =
f_{T_1}(t)P(T_2=t|T_1=t)}$, where $f_{T_1}(t) = e^{-t}$ is the marginal
distribution of coalescence times at the first (or second) locus and
${P(T_2=t|T_1=t) = e^{-\rho\lambda(t)}}$ is the probability of no change in
coalescence times given the coalescence time $t$ at the first locus. Here
$\lambda(t) = \frac{1}{4}\left(1-e^{-2 t}+2 t\right)$ is the exponential rate
of encountering a change in coalescence time along the chromosome given that
the local coalescence time is $t$ \citep{eriksson_sequential_2009,
carmi_renewal_2014}. Thus $R_0(t)$ is given by
\begin{equation}
	R_0(t) = e^{-t}e^{-\rho\lambda(t)}.
	\label{eq:r0t}
\end{equation}
This completes the solution of $R_j(t)$. Using Figure~\ref{fig:smcprimediagram},

\begin{equation}
	R^+(t) = I(t) + \sum_{j=1}^\infty R_j(t),
	\label{eq:rplussmcprime}
\end{equation}
where $I(t)$ is the probability that the process is in state $\bs{I}$ at time
$t$. Using \eqref{eq:smcprimesolution1} and \eqref{eq:r0t} we get

\begin{align}
	\begin{split}
		\sum_{j=1}^\infty R_j(t) &= R_0(t)\sum_{j=1}^\infty\frac{\left[\frac{\rho}{2}(1-e^{-2t})\right]^j}{j!} \\
									   &= e^{-t}e^{-\frac{\rho}{4}(1+2t-e^{-2t})}\left[e^{\frac{\rho}{2}\left(1-e^{-2t}\right)}-1\right].
	\label{eq:rjsum}
	\end{split}
\end{align}
Next, $I(t)$ satisfies the forward Kolmogorov equation

\begin{equation}
	I'(t) = \sum_{j=2}^\infty (j-1)R_j(t) - 2I(t),
\end{equation}
the solution to which is 

\begin{align}
	\begin{split}
		I(t) &= e^{-2t}\int_0^t e^{2u}\sum_{j=2}^\infty (j-1) R_j(u)du\\
			 &= e^{-2t}\int_0^t
		R_0(u)\left\{2e^{2u}+e^{\frac{\rho}{2}\left(1-e^{-2u}\right)}\left[(\rho-2) e^{2u}-\rho\right]\right\}du\\
	&= e^{-2 t} \bigg\{1-e^{\frac{1}{4} \left(-2 t (\rho-2)+\rho
-e^{-2 t} \rho \right)} \\
& \hspace{1 cm} -e^{-\frac{\rho}{4}}2^{\frac{\rho-4}{2}} (-\rho)^{-\frac{\rho-2}{4}} \left[\Gamma\left(\frac{\rho-2}{4} ,-\frac{\rho
}{4}\right)-\Gamma\left(\frac{\rho-2}{4},-\frac{e^{-2 t} \rho}{4}  \right)\right]\bigg\}.
	\end{split}
	\label{eq:probI}
\end{align}
Here, $\Gamma(a, b) = \int_b^\infty x^{a-1}e^{-x}dx$ is the incomplete gamma
function.

Together \eqref{eq:r0t}, \eqref{eq:rplussmcprime}, \eqref{eq:rjsum}, and
\eqref{eq:probI} give the joint distribution \eqref{eq:jointdistribution} for
the SMC'. For the ARG and SMC, the quantities analogous to $R_0(t)$ and
$R^+(t)$ for these models can be obtained by exponentiating the rate matrices
implicit in Figures~\ref{fig:SMCdiagram} and \ref{fig:ARGdiagram}. For the SMC,
the joint distribution can also be derived using the representation
\eqref{eq:smcrepresentation}.

The walk on the states $\bs{R_0}, \bs{R_1}, \bs{R_2}, \dots$ constitutes a
birth-death process with killing, where birth events correspond to additional
recombination events taking the process from $\bs{R_i}$ to $\bs{R_{i+1}}$,
death events correspond to coalescence events that take the process from
$\bs{R_i}$ to $\bs{R_{i-1}}$, and killing events, which take the process to an
absorbing state, here correspond to coalescence events that take the process to
$\bs{C_L}$, $\bs{C_R}$, or $\bs{I}$.  Under this formulation, the birth rate is
constant $\lambda_i = \rho$, the death rate is linear $\mu_i = i$, and the
killing rate is linear $\gamma_i = i+1$. This class of processes was studied by
\citet{van_doorn_birth-death_2005}, who demonstrated a different approach for
calculating $R_i(t)$. This alternative approach (not shown) confirms our
derivation of \eqref{eq:r0t}.

\subsection{Calculating asymptotic bias}

We are interested in estimating a single relative population size $\eta$
(defined relative to the true population size, $N$), which must be incorporated
into the transition density function $q(t|s)$ at recombination sites under the
SMC and SMC'. Under the SMC, this transition density function is

\begin{equation}
	q_{\textrm{SMC}}(t|s;\eta) = 
	\begin{cases}
		\frac{1}{s}(1-e^{-t/\eta}) & t < s \\[1em]
		\frac{1}{s}e^{-(t-s)/\eta}(1-e^{-s/\eta}) & t > s.
	\end{cases}
	\label{eq:qtssmc}
\end{equation}
This is equivalent to the conditional density
\eqref{eq:conditionaldistributionSTSMC} with the addition of a relative
population size parameter. Under the SMC', the transition function is

\begin{equation}
	q_{{\rm SMC'}}(t|s;\eta) = 
		\begin{cases}
			\frac{\frac{2}{\eta}\left(1-e^{-2t/\eta}\right)}{1+\frac{2s}{\eta}-e^{-2s}} & t < s \\[1em]
			\frac{\frac{2}{\eta}e^{-(t-s)/\eta}\left(1-e^{-2s/\eta}\right)}{1+\frac{2s}{\eta}-e^{-2s}} & t < s,
		\end{cases}
	\label{eq:qtssmcprime}
\end{equation}
which is equivalent to the conditional density
\eqref{eq:conditionaldistributionSTSMCPrime} with a relative population size parameter
included.

Under the SMC, given the local coalescence time $t$, the distance along the
chromosome until the nearest recombination event (measured in units of $\rho$)
is exponentially distributed with rate $t$ \citep{mcvean_approximating_2005}.
The likelihood function for a single relative population size $\eta$ under the
SMC is thus

\begin{align}
	\begin{split}
		L_{{\rm SMC}}(\eta | \{(t_i, l_i)\}) &= \frac{1}{\eta}
		e^{-\frac{t_1}{\eta}} \prod_{i=2}^k q_{{\rm
		SMC}}(t_{i}|t_{i-1};\eta) \prod_{i=1}^k t_i e^{-t_i l_i}\\
		&\propto \frac{1}{\eta} e^{-\frac{t_1}{\eta}} \prod_{i=2}^k q_{{\rm SMC}}(t_{i}|t_{i-1};\eta).
	\end{split}
	\label{eq:smclikelihood}
\end{align}
Under the SMC', the likelihood function for a relative population size
$\eta$ is

\begin{equation}
	L_{{\rm SMC'}}(\eta | \{(t_i, l_i)\}) = \frac{1}{\eta} e^{-\frac{t_1}{\eta}} \prod_{i=2}^k q_{{\rm
	SMC'}}(t_{i}|t_{i-1};\eta) \prod_{i=1}^k \lambda(t_i,\eta)
	e^{-\lambda(t_i,\eta) l_i},
	\label{eq:smcprimelikelihood}
\end{equation}
where $\lambda(t, \eta) = \frac{1}{4}\left[\eta(1-e^{-2 t/\eta})+2 t\right]$ is
the exponential rate of encountering recombination events that change the
coalescence time when the local coalescence time is $t$
\citep{eriksson_sequential_2009}. Note that under the SMC, the length $l_i$ of
a segment is independent of the relative population size $\eta$ given the local
coalescence time $t_i$. This is not true for the SMC', since the probability
that the coalescence time changes at a recombination site depends on the
population size.

As the length of the chromosome increases and the number of coalescence-time
changes goes to infinity, the asymptotic maximum-likelihood estimate
$\hat{\eta}^*$ of the relative population size under the SMC is

\begin{align}
	\begin{split}
		\hat{\eta}^* &= \lim_{k \to \infty} \argmax_\eta\ \frac{1}{\eta}
		e^{-\frac{t_1}{\eta}}
		\prod_{i=2}^k q_{{\rm SMC}}(t_{i}|t_{i-1};\eta) \\
	&= \lim_{k \to \infty} \argmax_\eta\ \left\{\log\left(\frac{1}{\eta}
		e^{-\frac{t_1}{\eta}}\right) + \sum_{i = 2}^k \log\left[q_{{\rm
	SMC}}(t_{i}|t_{i-1};\eta)\right] \right\}\\
		     &= \lim_{k \to \infty} \argmax_\eta\ \sum_{i=2}^k \log\left[q_{{\rm SMC}}(t_{i}|t_{i-1};\eta)\right] \\
			 &= \argmax_\eta\ \E_{{\rm ARG}}\big[\log(q_{{\rm SMC}}(T|S;\eta))\big] \\
			 &= \argmax_\eta\ \int_0^\infty \int_0^\infty \pi_{{\rm SMC'}}(s)
			    q_{{\rm SMC'}}(t|s;1) \log\left(q_{{\rm SMC}}(t|s;\eta)\right) dtds \\
		&\approx 0.95.
	\end{split}
	\label{eq:biascalculation}
\end{align}

Here the penultimate equality holds only if there is ergodic (i.e.,
law-of-large-numbers-like) convergence of the sequence of pairs of coalescence
times on either side of a recombination site under the ARG. In the Supporting
Information, we show that the continuous-chromosome pairwise ARG is ergodic.
That is, the mean coalescence time across a long chromosome converges to the
mean coalescence time at a single point along the chromosome. We are unable to
prove the ergodicity of the sequence of pairs of coalescence times at
recombination sites where the coalescence time changes; instead, we note that
\eqref{eq:biascalculation} is supported by simulation (see above). We also note
that \citet{wiuf_consistency_2006} proved the ergodicity of the discrete-locus
ARG under a variety of neutral demographic models. A similarly in-depth proof
may also apply for continuous-chromosome models, but we do not explore the
point further.

In \eqref{eq:biascalculation}, the ultimate equality follows from the fact that
the joint distribution of coalescence times is marginally the same at
recombination sites under the ARG and the SMC'. Numerical maximization of the
double integral shows that the maximum-likelihood estimate of a single
population size $N$ under the pairwise SMC is asymptotically biased, with the
asymptotic estimate being approximately $0.95N$.

Under the ARG, the stationary distribution of lengths between recombination
events that change the local coalescence time (i.e., the identity-by-descent
segment length distribution) is slightly different from that of the SMC'. (They
are different because subsequent recombination events ``heal'' with slightly
different probabilities under the ARG, while under the SMC', each subsequent
recombination event ``heals'' with the same probability.) Under the ARG, the
identity-by-descent (IBD) length distribution is not currently known. Given
that under the SMC' the maximum-likelihood estimator for a relative population
size involves the observed lengths, it is not currently possible to calculate
the asymptotic bias of the pairwise SMC' maximum-likelihood estimator of a
single population size. However, the IBD length distribution under the ARG is
approximated very closely by the SMC' IBD length distribution
\citep{carmi_renewal_2014}, so the SMC' estimator is likely to be nearly
asymptotically unbiased.

\pagebreak
\setcounter{equation}{0}
\setcounter{section}{0}
\setcounter{figure}{0}
\setcounter{table}{0}
\setcounter{page}{1}
\makeatletter
\renewcommand{\theequation}{S\arabic{equation}}
\renewcommand{\thefigure}{S\arabic{figure}}
\renewcommand{\bibnumfmt}[1]{[S#1]}
\renewcommand{\citenumfont}[1]{S#1}
\section{Supporting Information}

\subsection{Coalescence time distributions at recombination sites}

Here, we derive the joint, marginal (i.e., one-locus), and conditional
distributions of coalescence times at recombination sites where the coalescence
time changes under the ARG, SMC', and SMC. The distributions related to the ARG
and SMC' are derived from analysis of the continuous-time Markov chains
representing coalescence times at such recombination sites under these models
(Figure~\ref{fig:ARGSMCPrimeConditional}). Under the ARG and SMC', the joint
density function of coalescence times at recombination sites that change the
coalescence time (i.e., the joint density of $S$ and $T$) is 

\begin{equation}
	f_{S,T}(s,t) = 
	\begin{cases}
		\frac{3}{4}\left(1-e^{-2s}\right)e^{-t} & s < t \\[0.75em]
		\frac{3}{4}\left(1-e^{-2t}\right)e^{-s} & s > t,
	\end{cases}
	\label{eq:jointdistributionSTSMCPrime}
\end{equation}
and the marginal density function of $S$ (or $T$) is

\begin{equation}
	\pi(s) = \frac{3}{8}e^{-s}\left(2s+1-e^{-2s}\right).
	\label{eq:pisSMCPrime}
\end{equation}
The conditional distribution of $T$ given $S$ is

\begin{equation}
	f_{T|S}(t|s) = \frac{f_{S,T}(s,t)}{\pi(s)} = 
	\begin{cases}	
		\frac{2\left(1- e^{-2 t}\right)}{1-e^{-2 s}+2 s}  & t < s \\[0.75em]
  \frac{2 e^{-(t-s)} \left(1-e^{-2s}\right)}{1-e^{-2 s}+2 s} & t > s.
	\end{cases}
	\label{eq:conditionaldistributionSTSMCPrime}
\end{equation}

Equations \eqref{eq:jointdistributionSTSMCPrime}, \eqref{eq:pisSMCPrime}, and
\eqref{eq:conditionaldistributionSTSMCPrime} hold marginally at recombination sites
where the coalescence time changes under both the ARG and SMC'. Equations
\eqref{eq:pisSMCPrime} and \eqref{eq:conditionaldistributionSTSMCPrime} were
derived for the SMC' by \citet[][see eqns.\ (8) and (9),
respectively]{carmi_renewal_2014}, confirming our derivation.

Under the SMC the process for generating coalescence times at recombination
sites is equivalent to the continuous-time Markov chain in
Figure~\ref{fig:ARGSMCPrimeConditional}B with the transition rates from
$\bs{R_1}$ to $\bs{C_L}$ and $\bs{C_R}$ equal to $1$ instead of $3/2$. Under
this model for the SMC, the joint density of coalescence times on either side
of a recombination event is 

\begin{equation}
	f_{S,T}(s,t) = 
	\begin{cases}
		e^{-t}(1-e^{-s}) & s < t \\[0.75em]
		e^{-s}(1-e^{-t}) & s > t
	\end{cases}
	\label{eq:jointdistributionSTSMC}
\end{equation}
and the marginal density of $S$ (or $T$) is 

\begin{equation}
	\pi(s) = s e^{-s}.
	\label{eq:pisSMC}
\end{equation}
The conditional distribution of $T$ given $S$ under the SMC is 

\begin{equation}
	f_{T|S}(t|s) = \frac{f_{S,T}(s,t)}{\pi(s)} = 
	\begin{cases}	
		\frac{1-e^{-t}}{s} & t < s\\
	\frac{e^{-(t-s)}(1-e^{-s})}{s} & t > s,
	\end{cases}
	\label{eq:conditionaldistributionSTSMC}
\end{equation}
which confirms the derivation of \citet[][cf. their Eq.~(S6)]{li_inference_2011}.

\subsection{Pairwise ARG is ergodic}

Here we show that the pairwise ARG is sequentially ergodic. Let $\{t(x)\}_{x \ge
0}$ represent the random pairwise coalescence time at point $x$ along two
aligned, continuous, infinitely-long chromosomes modeled by the ARG.  Let time
be scaled such that the marginal distribution of $t(x)$ is exponential with rate
$1$ for all $x \ge 0$, and thus $\E[t(x)] = 1$.  Let the distance across the
chromosome be measured such that a segment of length $l$ recombines apart back
in time at rate $l/2$.  (Equivalently, a recombination event happens in the
chromosome interval $(x, x+dx)$ in the time interval $(t, t+dt)$ with
infinitesimal probability $dx\,dt$.)

One useful property of $t(x)$ is that it is strongly stationary. That is, the
joint distribution of $\{t(x)\}_{a \le x \le b}$ is the same as the joint
distribution of $\{t(x)\}_{a+h \le x \le b+h}$ for all $0\le a<b$ and $h>0$. To see
that this is the case, consider the \citet{wiuf_recombination_1999} algorithm for
constructing an ARG sequentially across the chromosome: at a given point, a
genealogy is drawn from the marginal distribution of genealogies, and then the
algorithm proceeds along the chromosome generating recombination events and
genealogies, where at each point along the chromosome, such events are drawn
from the conditional distribution given all previous coalescence and
recombination events. The initial point from which the marginal genealogy is
drawn has no effect on the resulting joint distribution of genealogies.

A stationary process $t(x)$ is ergodic if the covariance function $r(x)$
converges to zero as $x$ goes to infinity \citep{karlin_first_1975}.  Under the
ARG, the covariance function is

\begin{equation}
	r(x) = \frac{x+18}{x^2+13x+18},
\end{equation}
which satisfies this condition. Thus the pairwise ARG is sequentially ergodic: 
the mean coalescence time across a long chromosome converges to the mean
coalescence time at a single point. A similar proof could be given for the
discrete-locus ARG with evenly spaced loci, which has a covariance function of
the same form as the continuous-chromosome ARG.

\subsection{Supplementary Figures}

\begin{figure}[h!]
	\centering
	\includegraphics[width=0.4\textwidth]{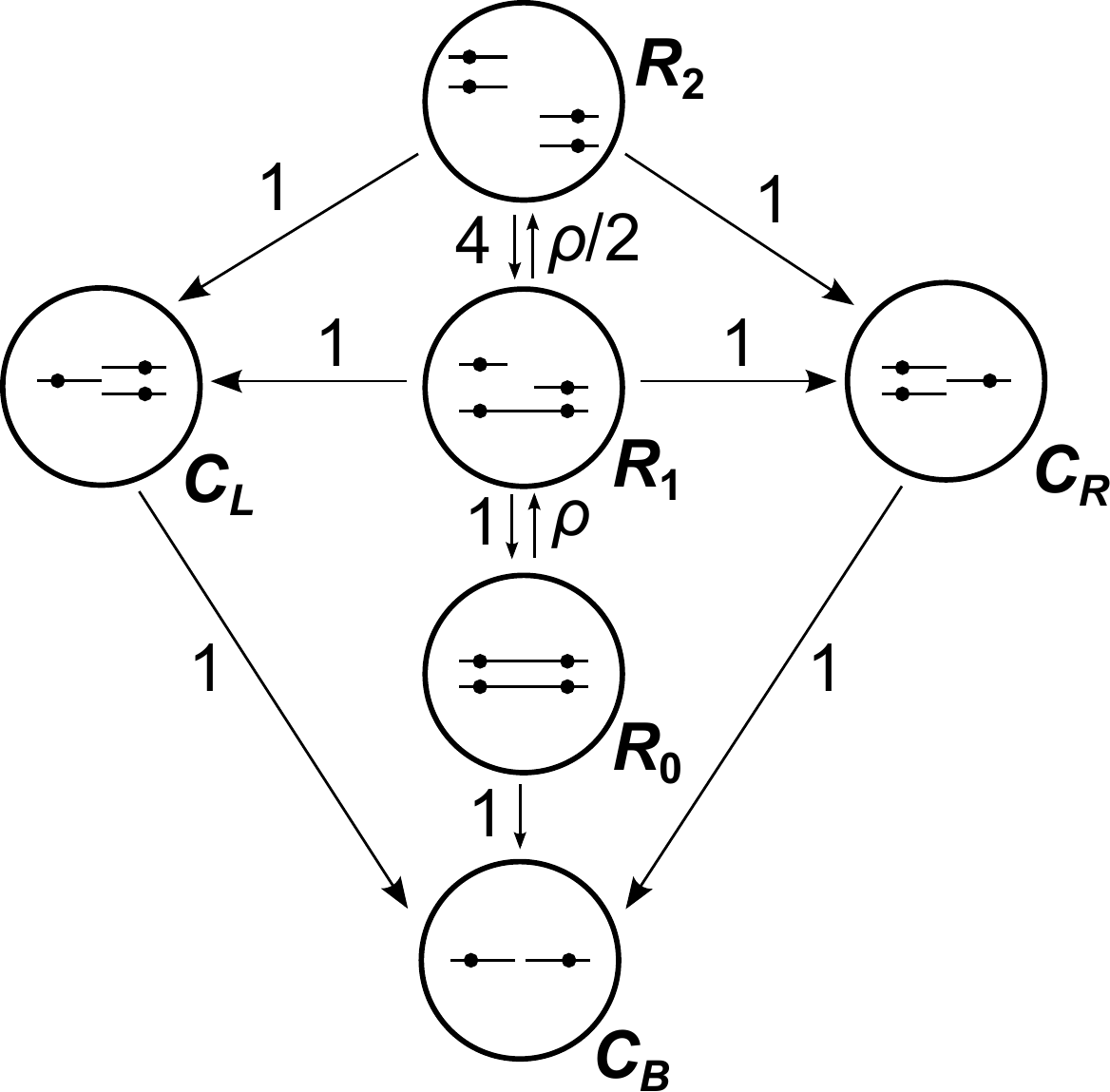}
	\caption{Schematic of the ARG back-in-time Markov process for two loci. The
	process starts in state $\bs{R_0}$, and transitions to other states occur
	with the rates indicated by arrows between states.} 
\label{fig:ARGdiagram}
\end{figure}

\begin{figure}[h!]
	\centering
	\includegraphics[width=0.5\textwidth]{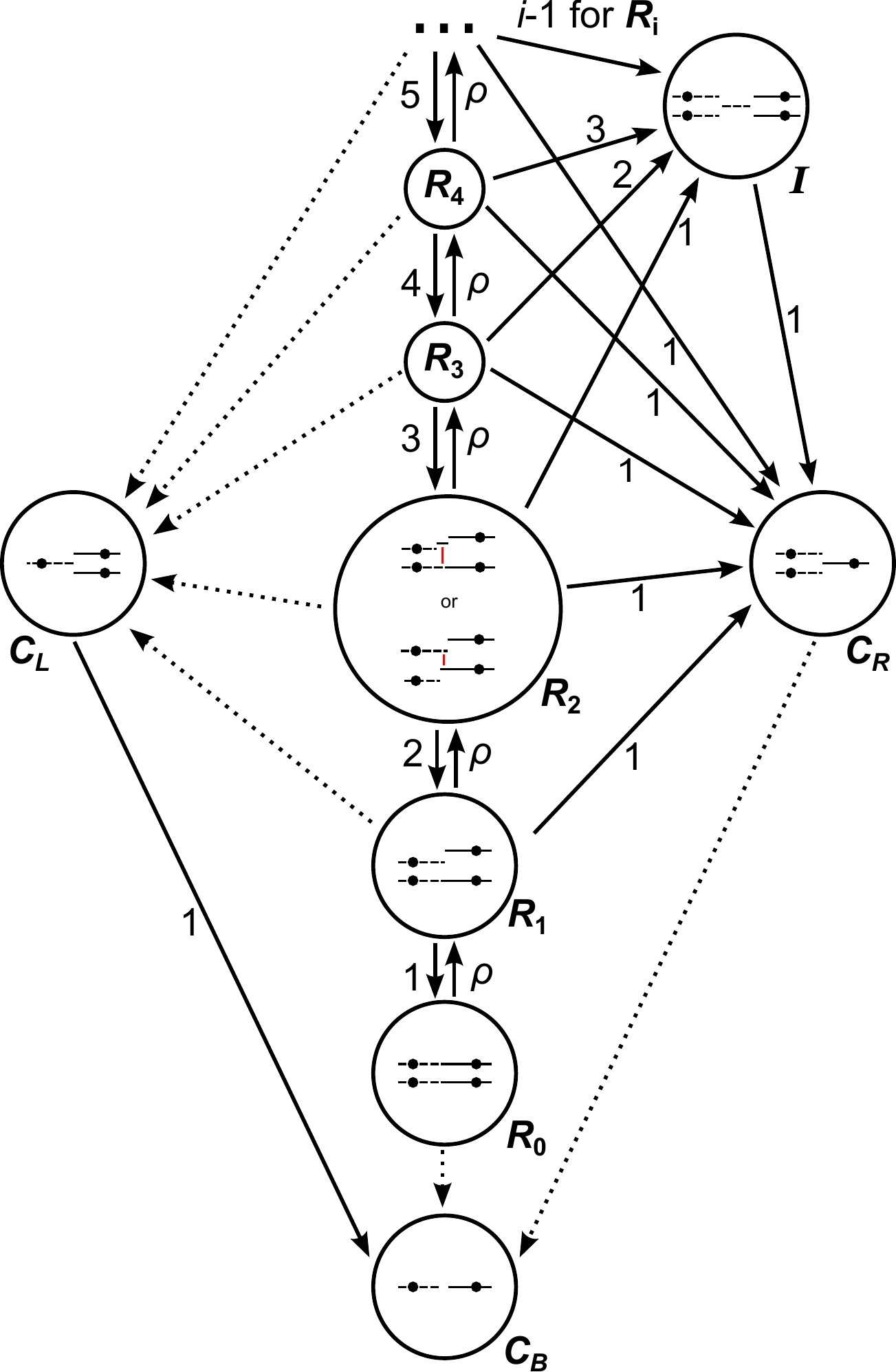}
	\caption{Back-in-time Markov process for generating a coalescence time
	$T_2$ at the right locus conditional on the time $T_1 = t_1$ at the left locus under
	the SMC'.  Starting at time zero in state $\bs{R_0}$, the process
	follows the transitions indicated by the solid arrows at the rates
	accompanying these arrows.  Transitions indicated by dotted
	arrows are followed instantaneously at time $t_1$. See
	\citet{hobolth_markovian_2014} for analogous processes for the ARG and SMC
	models.}
\label{fig:smcprime_conditional_diagram}
\end{figure}

\end{document}